\documentclass[pre,prl,twocolumn,groupedaddress]{revtex4-1}

\usepackage{graphicx}
\usepackage{amssymb}
\usepackage{subfigure}

\begin{document}


\title{A model for dynamical evolution of science in space}


\date{July 29, 2014}
\author{Jan Moritz Joseph}
\author{Jens Christian Claussen}\thanks{Corresponding author.}
\email[]{j.claussen@jacobs-university.de}
\affiliation{Computational Systems Biology Lab, Jacobs University Bremen, Campus Ring 1, D-28759 Bremen, Germany}


\date{July 30, 2014}

\begin{abstract}
How does the topological space of science emerge? Inspired by the concept of 
maps of science, i.e.\ mapping scientific topics to a scientific space,
we ask which topological structure a dynamical process of authors collaborating and publishing papers can generate. We propose a dynamical process where papers as well as new groups receive topical positions embedded in a
two-dimensional euclidean space. The precise position of new papers depends on previous topics of the respective authors and is chosen randomly in a surrounding neighborhood including novelty and interdisciplinarity. Depending on parameters, the spatial structure resembles a simple Gaussian distribution, or spatial clusters of side-topics are observed. We quantify the time-evolution of the spatial structure and discuss the influence of inhomogenities.
\end{abstract}


\maketitle

\section{Introduction}
\noindent
{\sl ``What is knowledge?''\\[-1ex] \mbox{} \hfill ({\sc Plato}, 
$\Theta\varepsilon\alpha\acute{\iota}\tau\eta\tau o\varsigma$%
%
)}
\\[.5ex]
\indent
Science is a process of generating knowledge.
Know\-ledge is generated by scientists, performing and analyzing experiments,
structuring knowledge by models and stating predictions, 
and drawing conclusions through mathematical reasoning.
Scientists of all disciplines act, interact and trade their knowledge at universities,
which themselves evolve largely self-organized into structures of high
complexity in their social interactions and also with respect to their
scientific structure.
But how can the scientific structure of a university, or the world community of scientists,
be characterized -- and why does it emerge to obviously highly complex structures? 
These are questions which, if answered, would provide deeper understanding into the
dynamics of the scientific process itself, and the aim of this paper is to step into this direction.

But how can the complex structure of science be re\-pre\-sen\-ted in an appropriate space
and what does it look like?  
The most extensive studies in this direction have been provided by Katy B\"orner and coworkers
\cite{Boyack2005,boerner}
who analyzed large datasets of scientific papers across disciplines and embedded them
in a metric space according to neighborhood relations based on text similarity:
publications with the larger overlap in scientific vocabulary are assigned the closer
distance in scientific space.
Their work 
revealed
that the scientific structure of the investigated datasets
always showed strong clustering in the known classic scientific disciplines
with some overlap regions of interdisciplinary work. 
Across the disciplines, there was also evidence for local sub-clusters comprising a 
hierarchical and eventually fractal structure.
While -- for convenience of illustration -- restricting the embedding dimension to two,
these properties are largely preserved in higher-dimensional embedding; 
and the two-dimensional ``maps of science'' 
\cite{Boyack2005}
coarse-grain our level of ignorance by projection down to a 
two-dimensional space. 

According to a traded 
metaphor
{\sl 
``a university is just a group of buildings gathered around a library''}
\cite{regensburg},
and this is akin of a seed of any model of science:
Scientists provide knowledge to the library, interact through it, and
grow into the surrounding. 
Such a process centrally involves attachment of scientists and their newly generated 
scholarly work to preceding work and scientists,
which in several disciplines have been shown to exhibit scale-free properties
\cite{scalefree_arxiv,barabasialbert99},
which go beyond a small-world structure
\cite{wattsstrogatz},
as a consequence of a Matthews effect
\cite{barabasialbert99,simon53,bornholdt_ebel01}.

But how do human decisions -- on which topic to work and publish in science -- 
influence the generated network and, thereby, the organizatorial structure of science? 
To address such questions, a class of dynamical models is needed
grounding a framework for more detailed modeling and interpretation of data.

The aim of our model 
is to reproduce basic
stylized facts of the scientific process:
(i) the spatial structure is clustered with hierarchical substructures,
(ii) scientific papers and the collaboration network show scale-free properties
through richer-get-richer mechanisms, 
and
(iii) the spatial distributions of papers and authors in scientific space are strongly non-gaussian.
In this paper, we introduce a model that dynamically generates structures 
reproducing these stylized facts.
Contrary to our initial intuition, we had to include specific mechanisms into the
dynamics to avoid the model to collapse its spatial complexity 
(into a plain gaussian 
cloud of a consensus topic with some variations around):
We had to prevent scientists from growing their group and offsprings for eternity by
a realistic ``retirement'' process. 
And we included a preference bias for interdisciplinarity for
each two (or more) scientists initializing new work. 
We expected initially both mechanisms to influence the spatial growth, but
not to be a crucial component.
According to our study of possible model variants, 
however they seem to be fundamental 
to the evolution of scientific landscapes.


\section{The model}
 In our model,
we consider scientific collaboration networks containing $n$ productive authors. Two authors are considered 
to be connected 
if they have coauthored a publication. Here, authors that are 
 acquainted, 
yet have not collaborated, are also connected. The graph is weighted; plain acquaintanceship is equal to weight one and a weight of $k+1$ corresponds to $k$ collaborative publications. The authors are connected via initial connection in the collaboration graph via preferential attachment. After an author 
 publishes one paper
with a probability of 0.075, 
 a new connection 
 is established
to another author via preferential attachment.

A position $\vec{a_i}\in \mathbb{R}^m$ is assigned to each author in a scientific space, which has $m$ dimensions and is metric. Initially authors are randomly distributed following 
 a Gaussian
distribution. A randomly drawn 
{\sl interdisciplinary factor} or, synonymeously, {\sl strategic value} $s_i$
is assigned to each author. This interdisciplinarity factor depicts whether authors prefer collaborations and research in areas distant to its own position (interdisciplinary strategy) to research within its neighborhood (disciplinary strategy).  

Scientists produce papers on their own or via collaborations. The ratio of collaborative and non-collaborative publications 
is chosen as $3:1$, i.e.,
25 per cent of the papers are single author papers. Papers are situated in the scientific space along with authors:

\begin{figure}[htbp]
\includegraphics[width = 0.75\columnwidth]{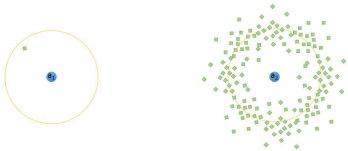}%
	\caption{Possible positions of papers (green) for single author (blue) publications as well as the position of a single publication in two dimensions: Papers are placed on 2-spheres around the authors position. The radius of the spheres is determined by the strategic values of the author. \label{fig:Paper_distribution_one_author}}
\end{figure}	
	
\begin{figure}[htbp]
\includegraphics[width = 0.65\columnwidth]{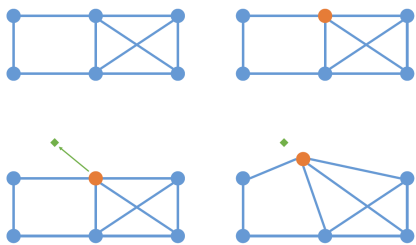}%
\caption{Publication process for a single author: Authors and their collaboration network is depicted in blue. One author is selected (orange) to publish a paper (green). The position of the author changes due to the publication. \label{fig:single_author_pictogram}}
\end{figure}

\begin{figure}[htbp]
\includegraphics[width = 0.75\columnwidth]{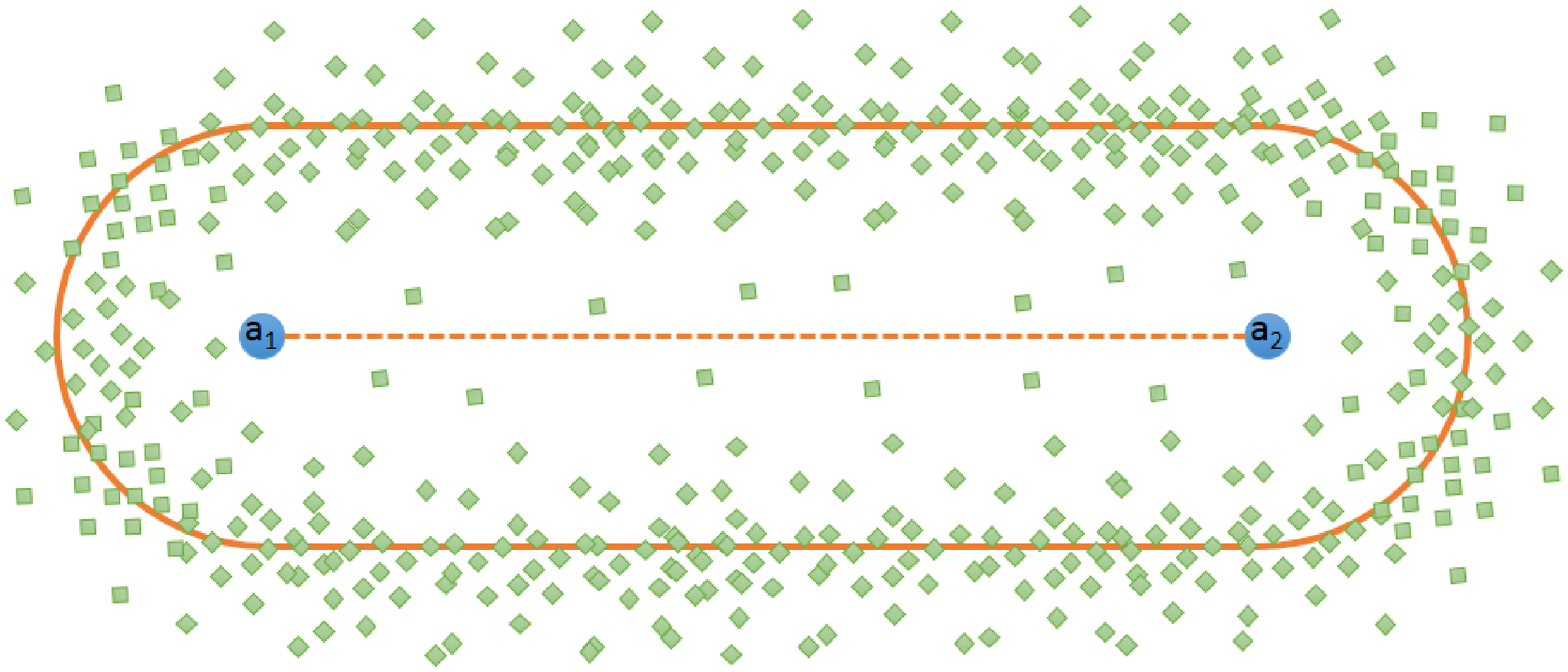}%
	\caption{Possible position of papers (green) for collaborative publications of two authors (blue) in two dimensions: The combined strategic values of the authors leads to a mean value of the Poisson distribution of distances for papers (orange). Papers are placed on 2-spheres around random positions on the connection line (orange, dotted) between the authors. Here strategic values were not associated with the authors. \label{fig:Paper_distribution_two_authors}}
\end{figure}

\begin{figure}[htbp]
\includegraphics[width = 0.95\columnwidth]{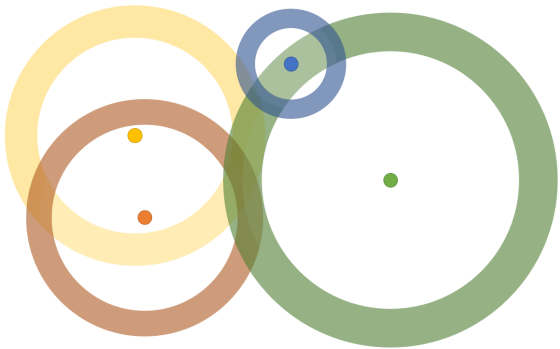}%
	\caption{Four authors in different colors are shown for two dimensions. Each authors has a corresponding strategic value. The strategic values are the mean values for a Poisson distribution that determines the radius for 2-spheres around the author's  position. The area in which each author prefers to publish is shown by "doughnuts" in corresponding colors. In consequence, the yellow and the green author are likely to collaborate since their areas of interest overlap. On the other hand, the red and the blue author are less likely to collaborate since their areas of interest do not overlap. \label{fig:publication_likelyhood}}
\end{figure}

\begin{figure}[htbp]
\includegraphics[width = 0.7\columnwidth]{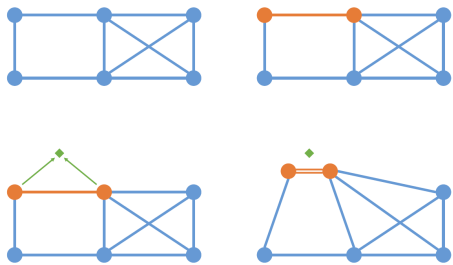}
\caption{Publication process for a collaborative paper: Authors and their collaboration network is depicted in blue. One collaboration and its corresponding authors are selected (orange) to publish a paper (green). The position of the authors changes due to the publication. \label{fig:collaborative_pictogram}}
\end{figure}

\begin{description}
	\item[single-author publication] The paper's position is located on a $m$-sphere around the author's position $\vec{a_i}$ (the author's index is $i$) as shown in Fig.~\ref{fig:Paper_distribution_one_author}. Positions on the spheres are drawn following a uniform distribution. However, the radius of each sphere is drawn from a Poisson distribution with variance equal to the interdisciplinary factor $s_i$ of the researcher. This yields
\begin{equation}
	\label{eq:author_position_shere}
	\vec{p}_j\in S^{m-1} = \{ \vec{x}\in\mathbb{R}^m: ||\vec{x}-\vec{a_i} ||_2=r_i \},
\end{equation}
Herein, $r_i$ is drawn with 
probability
\begin{equation}
	\label{eq:Poisson_distribution}
	\text{Pr}(r_i = k)= \frac{s_i^ke^{-s_i}}{k!}.
\end{equation}
	The probability for a publication 
 to be written 
is determined by the publication speed of the author. The fraction of the author's publication speed relative to the summed publication speeds is the probability to write a new paper. The process is depicted in fig.\ \ref{fig:single_author_pictogram}. 
	\item[collaborative publication] Collaborative papers are written by two authors as shown in 
fig.~\ref{fig:Paper_distribution_two_authors}. Both authors must be connected in the collaboration graph, thus they know each other or have published as co-authors. The paper is located on a $m$-sphere analog to the single author publication process. The center is located on the authors' connecting line and the radius is determined by the mean strategic value. A random, uniformly distributed location on that connecting line (following Euclidian norm) between the two authors is drawn. After the position is drawn, there is given a probability for a successful publication by the multiplied Poisson distributions with mean values fixed by the authors strategic values. The process is depicted in fig.~\ref{fig:publication_likelyhood}. 
	The probability for a collaborative paper is determined by the authors' publication speeds, their combined strategic values, and their distance: Fast-publishing authors continue at a high rate; authors with large distance collaborate if both have an interdisciplinary interest and vice versa. The process is depicted in fig.~\ref{fig:collaborative_pictogram}. After a collaborative publication the connection weight in the collaboration graph is increased by one making future collaborations more likely for the two scientists.
\end{description}
\vspace{1ex}
\clearpage

\begin{figure}[htbp]
\includegraphics[width=0.89\linewidth]{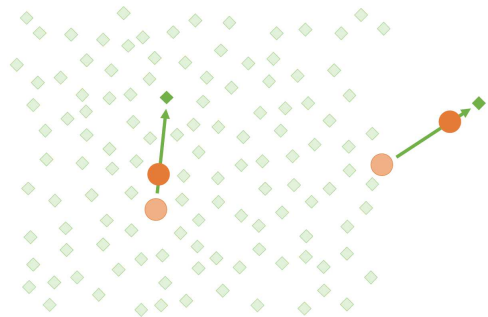}
	\caption{Two authors are depicted with their old position (light orange) and new position (dark orange) after writing a new paper (dark green). Older papers are shown in light green. The authors' position is varied stronger if fewer papers are written in the neighborhood of its new publication.\label{fig:novetly_reward_influence}}
\end{figure}

After a new paper is published, the corresponding author's positions are adopted. 
The new author's position $a_{\text{new}}$ is given by
\begin{equation}
	a_{\text{new}} = (1-n_f)a_{\text{old}}+n_f p_{\text{new}},
\end{equation}
where $a_{\text{old}}$ is the previous position of the author, $p_{\text{new}}$ is the paper's position and $n_f$ is the adjusted novelty reward 
mapped onto
 the interval $[0, 1]$ by applying a monod-hyperbolic-function (saturation function):
\begin{equation}
\label{eq:saturation_function}
	n_f = \frac{0.9 n_r}{1.75 + n_r}.
\end{equation} 
A novelty reward $n_r$ is given for each paper by the density of papers in the neighborhood; it is calculated as the mean distance between the nearest ten percent of all papers written and the paper itself. Authors adopt their position according to the rewards: If the density is low, the paper has a high novelty reward and thus the author vastly adopts its position and vice versa, as shown in fig.~\ref{fig:novetly_reward_influence}. 


%
%
\section{Results}
 To enter into realistic ranges of scientific (sub-) communities, we performed
extensive simulations in two dimensions with 500 authors that produced up to 24,000 papers.
The intervals for interdisciplinary factors as defined in the model were 
surveyed.
Furthermore authors with high publication numbers were 
separated from the model to investigate their influence. 

To describe the dynamics
 by means of macroscopic, order-parameter-like variables,
 the mean position of the papers' distribution, their standard deviation and the kurtosis were observed while 
 varying
 the interval size of the strategic vales from $[0.1, 0.3]$ to $[0.1, 2.5]$, 
to compare the emerging dynamics.
 Furthermore intervals without disciplinary authors were simulated. Here intervals from $[0.5, 0.7]$ to $[1.0, 1.9]$ were chosen. The distributions 
herein are homogeneous, i.e.,
assume that no strategy is preferred by researchers within the interval. 
Authors with large interdisciplinary values have a significant influence on the dynamics of the system, since 
the resulting scientific map extends to a larger area.

\subsection{Papers' Distributions}
Before calculating the kurtosis the data were preprocessed as follows: First, the papers' positions were scaled in terms of standard deviations. This was done via the Mahalanobis distance for each paper. It was calculated in every dimension and the sign of the original position was preserved. Second, the principal components of this distribution were calculated. The kurtosis was evaluated along these axes. If the system evolves isotropically, the kurtosis will have similar curve progression for each basis. However, if the distribution is not isotropic and thus there are large differences between the covariances along the principal components, this will be depicted in varying kurtosis curves. 
	
	During the time evolution of the model, single and outlying research groups evolve. These groups are drivers for a non-isotropic distribution of papers, which can be regularly observed for the majority of the simulations. In consequence, the kurtosis differs for a single instance of the model during a period of time. However, this evolution can be reversed by three processes: At first, the outlying groups are attracted to their collaborators, which in most cases are not outlying themselves. Secondly, the collaborators are reversely attracted to the outlying scientists. Thirdly, new outlying groups evolve. These groups only receive high novelty rewards, if and only if they are situated off the main and dominating principal components of the system. (If they are situated along a principal component, and are not on the edge of the system, the novelty reward will be small due to already published papers in that region.) Furthermore, groups with high novelty reward will move fast in the perimeters of the scientific space filled with research. 

As the interdisciplinarity parameter introduces a bias to write papers in unoccupied scientific space,
an obvious outcome of the process is a spatial growth, i.e., the spatial dimension of the 
emerged structure increases. 
Both the standard deviation and the maximum range increase with the interval size. (Naturally both characteristics show analog development since they are measuring the same macroscopic property.) This observation is manifested in the model itself: Authors with larger interdisciplinary values will write papers (both single authored-papers and collaborative works) that have a farther distance to their current position. In addition the deviation of the mean is larger for bigger intervals since the system's dynamics is stronger influenced by outliers. Recapitulatory, with more interdisciplinary authors the system's dynamics is more exploratory and larger areas of the scientific space will be filled with papers. 

The development of the kurtosis is highly interesting. The curve progression is similar in all models thus implying that the structure of the papers' distributions is similar in that manner. This is a remarkable result since interdisciplinary work enables research groups to explore areas completely free of previous research. These groups have high novelty rewards for their papers and mostly are attractive to collaborate with. Thus they are publishing at a high rate. In consequence, the scientific map is highly segregated. After preprocessing the papers' distributions to scale with standard deviation both isotropies and anisotropies are observed. These processes are main drivers for the scientific process: The research expands itself to new areas via anisotropies, which generates innovation. The isotropic processes consolidate the knowledge and enables detailed research in areas that are already initially filled with papers.

Thus the presence of interdisciplinary authors has a severe influence on the time evolution of the system. Without these groups new areas of the scientific space are not persistently researched. The distance between the established researchers and the outlining exploratory groups scales with the maximum range of the available strategies. Thus the density of the published papers is higher in the presence of groups working disciplinary in their field. 

\subsection{Social Contacts}
\begin{figure}[htbp]
\includegraphics[width=0.99\linewidth]{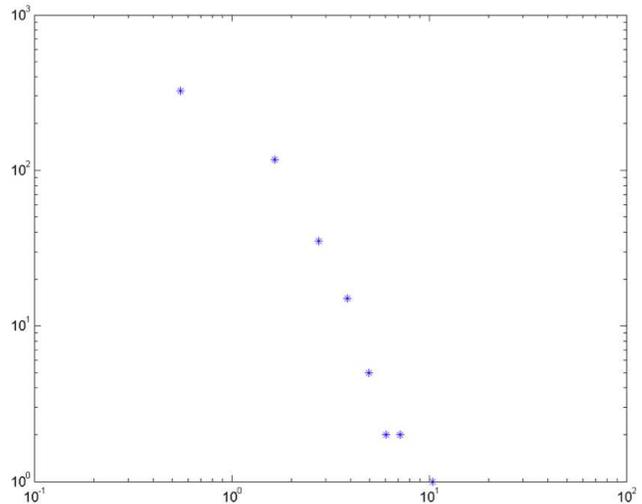}
	\caption{Collaboration graph with 500 authors after 24,000 simulated publications. The distribution follows power-law with exponential cut-off.}\label{fig:res:node_degree}
\end{figure}

 Next we verify the scale-free structure of the artificial collaboration graph generated from our model.
The distribution of the node degrees in a model with 500 authors is given as a double-logarithmic plot in 
fig.~\ref{fig:res:node_degree}. The figure shows a snap-shot after 24,000 paper were written. The degree distribution is given as power-law with exponential cut-off during the whole evolution of the system over time. 

The social graph in the scientific communities is examined in detail in \cite{NewmanPNAS2001} and \cite{NewmanPNAS2004}. Both the initial node degree distribution and the distribution during the time evolution of the model follow a power-law with exponential cut-off. 
Curve fitting on the model delivers analogous curves as 
as for figs. 1, 2 in \cite{NewmanPNAS2001}. 
Thus the model proposed in this paper complies very well with the properties of the scientific collaboration graph. In conclusion the selected simple model with preferential attachment performs very well, as the results comply with other research on real social networks.

\clearpage
\subsection{Inhomogeneity of Publications Speeds}
Both the distribution of authors' positions and the comparison of 
the models' 
publication speeds show that the inhomogeneity of the publication speed is vital for the development of an scientific map with nontrivial spatial structures.

\subsubsection{Authors' Positions \label{sec:res:authors_positions}}
During the time evolution of the model single outlying researchers emerge. 
These research groups are well-connected and publish at a high rate as shown in the following paragraphs. 
Outliers were detected using QQ-Plots. Here, the outliers were classified manually with the 95\%-quantile method.  Two different types of outliers are emerged: First, there are groups that are separated from the rest. Second there are dynamics such that the outliers are less distant and intermediates between the non-outlying and the strongly outlying groups are existent. Both cases were found in all simulations with strategies on an interval $[0.1, x], x\in[0.2, 2.5]$. Furthermore for each simulated interval, the number of outlying researchers was below 2\% of all researchers.

Single outlying groups emerge during the time evolution of the model. These researchers are well-connected and publish at a high rate. Their distance towards the covered scientific space grows with larger interdisciplinary factors. However, the distance from the mean measured in standard deviations does not grow fast or at all. 
\begin{figure}
\includegraphics[width=0.9\linewidth]{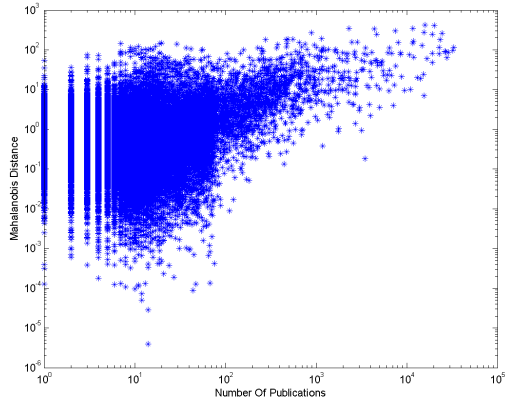}
\caption[distance vs. publication speed]{The double logarithmic scatter plot shows authors in all models in relation to their Mahalanobis distance to all scientists.} \label{fig:res:scatter_distance_publication_number_complete}
\end{figure}
The ten authors with the highest number of publications in the dynamics of each simulation are examined closer: These authors published in average each about 5\% off all published papers with a publication number distribution following a power-law. Furthermore these authors had a averaged Mahalanobis distance of 22 and a median Mahalanobis distance of 7.5 (which is equivalent to approx.\ 50 standard deviations). This relation is shown in detail in fig.~\ref{fig:res:scatter_distance_publication_number_complete}.

\begin{figure}[htbp]\includegraphics[width=0.9\linewidth]{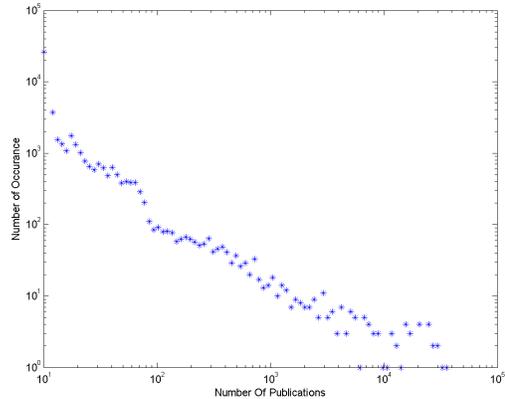}
	\caption[publication speed distribution]{Double-logarithmic plot of the binned distribution of publications speeds. The 100 bins have logarithmic scaled size from $10^1$ to $10^5$.} \label{fig:res:publication_speed_distribution}
\end{figure}

\vspace*{.8cm}
%
\subsubsection{Distribution of publication speeds}
The distribution of publications speeds follows a power-law distribution
as shown in fig.~\ref{fig:res:publication_speed_distribution}. 
Double-logarithmic binning is used to show the property. In the plot, 100 bins with logarithmic scaled size from $10^1$ to $10^5$ were used. This behavior of the system is to be expected since non-equilibrium distributions are the underlying statistics for many processes. 

A disparity between the amount of authors publications is vital for a segregating process. The major consequence of the disparity is the emergence of single, well-connected research groups publishing at a high frequency. These groups are often found in the outer areas of the covered scientific space. This is a self-enforcing process since a large record of publications attracts other scientists leading to new collaborative papers, which is an intrinsic dynamics in science. 

With strict deletion of highly-publishing authors the disparity is reduced. In consequence a new researcher filling a vacant position cannot hold its outlying location and is attracted back into the mean of the authors' positions on the map. Thus the process results in a Gaussian-shaped paper distribution. Regarding the distribution of the papers the process typically evolves with a small kurtosis and standard deviation.

\section{Discussion}
Two important drivers of the scientific process were observed. At first, there is inhomogeneity of publications speeds. 
Secondly, if not retiring fast-publishing authors,
 the generation of nontrivial spatial structure is prohibited, 
and the evolving scientific map collapses to a Gaussian bell without outstanding properties. 
Conversely, our model results in a dynamics similar to real scientific process as given in the well-researched "Map of Science" if authors do not retire and thus the evolution of vastly different publication speeds is possible. 
A fundamental property of the model is the non-equality distribution of publication speeds.
Apart from the higher publication speeds of outlying authors, another striking property of the model is 
that the scientific process is a superposition of both anisotropic and isotropic dynamics. On the one hand, anisotropy is driven by the development of innovative and new research. Authors, situated at the outer areas of the covered scientific space, have high novelty rewards for their research and thus cover large areas with publications at a low density. Naturally, authors with large interdisciplinary strategies are more exploring. Thus, the anisotropy is strongly influenced by these scientists. 
Likewise, if larger interdisciplinary strategies are available, the scientific map is more expanded.
On the other hand, the isotropic process extends the knowledge in areas of the scientific space that are already researched. The aforementioned outlying authors attract other research groups. Thus the outer areas are connected with the scientific process and the density of publications is increased. This process is slow in speed since the novelty rewards of those publications are low. This process is naturally induced by authors with lower interdisciplinary factors and the process evolves from groups of authors and not single research groups. This has an important consequence: While larger interdisciplinary strategies tend to extend
the spatial outreach of
 the scientific process, the structure of the process does not 
crucially depend on
this parameter. Scaled on standard deviations, papers' distributions are not depending on 
available fraction of 
authors pursuing
 interdisciplinary strategies.

\section{Conclusions and Outlook}
In summary, we have shown that the emergence of a complex structured topology
of scientific knowledge can be explained from a considerably simple 
dynamical model of authors writing papers.
 One key ingredient that we included was a bias towards interdisciplinary work or novel fields,
as authors' curiosity is assumed to be a natural part of the process.
 Unexpectedly, both an inhomogeneous publication speed had to be introduced,
as well as a retirement mechanism. 
 While we cannot exclude that other comparably simple models could be 
 set up without these mechanisms, 
 several obvious variants  that we considered only revealed a plain Gaussian distribution
of papers around a mainstream topic.
 Obviously, the scientific process in reality is able to go beyond, 
and structures itself into disciplines and topical clusters eventually exploring new 
subjects and interdisciplinary areas.

Of course, the model studied here was merely intended to describe the 
main stylized facts of such a process, and does not include finer details
that would be necessary to compare precisely to real data.
The framework we introduced offers ample flexibility to include realistic
details, such as 
including inhomogeneities in author preferences, especially preferences in the
 number of authors per paper, which may be ranging from 1 in large areas of pure mathematics
to thousands of authors in large-scale particle collider collaborations.
Apart from including more specific properties, 
as the refereeing process
\cite{thurner},
 an important application of such models is that 
 steering mechanisms of scientific management, be it on the political layer,
be it in university leadership, or simple in the mind of each research group leader,
can be implemented and their consequences on the topological structure 
as well as its local growth and impact can be analyzed.

As increasingly larger part of scientific research 
currently slave to funding mechanisms, ranking and reward mechanisms,
which themselves eventually depend on 
oversimplified, abstract, and content-ignoring 
measures as cumulatively summed journal impact factors or
citation metrics \cite{hirsch},
the precise influence of such modifications of the scientific process should be
studied thoroughly.
From this, researchers could gain insight to which fraction their scientific
strategies are based on free curiosity and to which extent science gets steered by 
external mechanisms.


%

\end{document}